# Towards a class of complex networks models for conflict dynamics

Vinicio Pelino[1,2], Filippo Maimone[2]

[1] Osservatorio per la Sicurezza Nazionale, Rome, Italy

[2] Italian Air Force, CNMCA, Pratica di Mare (Rome), Italy

Corresponding author: V.Pelino (pelino@meteoam.it)

## Abstract

*Using properties of isospectral flows we introduce a class of equations useful to represent signed complex networks free continuous time evolution Jammed and balanced states are obtained introducing a class of link potentials breaking isospectral invariance of the network. Applications to conflict dynamics in social and international relations networks are discussed*

## I INTRODUCTION

In this paper we are interested into antagonistic interaction dynamics inside a group of actors or, technically speaking, inside a signed complex network where relationships are always represented as mutual. One of the most interesting result of social psychology, the Heider's balance theory [1], states that people make every effort for cognitive balance in their network of likes and dislikes. Furthermore, social scientists have always empirically observed a triad closure effect, where new friendships tend to develop among people who currently have friends in common, therefore this theory is most easily understood in triadic relationships. Suppose we have a group of three actors $s_i$ $(i=1,2,3)$ and each pair has a distinct social link $A_{ij} = A_{ji}$ labeled $\pm 1$, denoting positive or negative sentiment. A state of the triad can be described by a triangle with edges representing interaction between persons (nodes). Up to node permutations, four possible configurations come out, with two triads having triple product $A_{ik}A_{kl}A_{lj} = -1$. Heider argued that, from a social-psychological perspective, negative triads have to be considered unstable or imbalanced. Balance theory asserts that balanced states (identified as positive product triads) will be preferred over imbalanced ones, and imbalanced states will lead to activities to change them into balanced states. For a generic graph $G=(V,E)$, where $V=\{s_i\}, (i=1,...,n)$ is the set of $n$ nodes and $E=\{\{s_i,s_j\}\}$ is the set of links, and adjacency matrix $A$, with $Tr(A)=0$, Cartwright and Harary demonstrated in 1956 the following *Structural Balance Theorem* :

"*If a labeled complete graph is balanced, then either all pairs of nodes are friends, or else the nodes can be divided into two groups, **A** and **B**, such that each pair of people in **A** likes each other, each pair of people in **B** likes each other, and everyone in **A** is the enemy of everyone in **B***" [2]

From a dynamical system viewpoint, balance theory is a static one. It tells us that, once balanced, the end state of a social group with positive and negative relations is a bipolar net of two warring factions; otherwise the whole group is cohesive, but the way it is constrained to reach these stable points is unknown to us. We will provide for a general framework to construct such a dynamics, introducing into the picture a minimal set of elements relevant to conflicts.

In Sec. II we give some preliminaries on recent results on dynamical models used to reach balance states. Sec. III is devoted to a discussion of walks on signed networks and their meaning using communicability matrix formalism. Isospectral flow and its related graph invariants is introduced in Sec. IV. Single and collective link potentials, and transitions to jammed and balanced final states according to Cartwright and Harary theorem, are discussed in SecV. A brief discussion on triadic relationship is provided in Sec.VI, while a more general class of models is discussed in Sec.VII. Finally, we provide some concluding remarks in Sec. VIII.

## II    PRELIMINARIES

In order to discover a dynamical theory for signed networks, first approaches were tempted using discrete time models [3] and interesting outputs of these simulation were unbalanced final networks called jammed states. They can be understood as local minima of the energy landscape of the system driven by the network potential energy:

$$u(A) = -\frac{1}{\binom{n}{3}} Tr(A^3) \qquad (1)$$

In other words function (1) represents the difference between numbers of balanced and unbalanced triangles $\Delta^{\pm}$ over the total number of length three closed paths : $u(A) = -\frac{\sum \Delta^+ - \sum \Delta^-}{\sum \Delta^{\pm}}$. Range extremes of $u(A) = \pm 1$ correspond, respectively, to totally unbalanced (+1) and balanced (-1) states. An important result of [4] is that jammed states $A_{jammed}$ cannot have energy above zero, $u(A_{jammed}) \leq 0,$ while final states assume the topological network configuration of Paley graphs.

Kulakowski et al [5] started to study Heider balance in a continuous time dynamics. They simulated a simple model for a single triad using general real numbers for the adjacency matrix to describe the opinion distribution and its dynamics. For larger networks, their method of inspection ceases to be simple, and therefore they relied on numerical simulations.

More recently, Marvel et al.[6] extended the analytical study to $n \times n$ continuous time dynamics. They proposed a dynamical system governing the time-evolution of social relationships assuming a symmetric matrix $X(t)$ whose off-diagonal terms represent the network adjacency matrix. In this way, they showed that the system reaches a balanced pattern of edges in a finite time by the Riccati matrix equation

$$\frac{dX}{dt} = X^2 \qquad (2)$$

for essentially any initial $X(0)$. In equation (2) networks naturally adjust themselves to a final global balance as described in the works of Heider; no jammed states are reached. We point out that even though nets reach the balanced state, we loose any information on the strength of enemy/friendship in a finite time because weights of links $X_{ij}(t) \to \pm\infty$, so that only link signs can make sense. Tests on graphs associated to real data have however shown that all large social networks live in an unbalanced state where friend and foe relationships continuously change with time [7], [8]. This fact gives us the opportunity to look for analogues of jammed states in continuous time dynamics. In this paper we will focus on the dynamics of undirected connected graphs of weighted adjacency matrices $A$, where self-loops are generally allowed. They will correspond, according to Ref. [6], to positive/negative self-trust of the actor $s_i$ with respect to the global network behavior (in a geopolitical context, considering a network approach to international relations [9], $A_{ii}$ may correspond for example to the population's feedback to the foreign policy of its own Government in choosing allies or enemies). These objects belong to the more general space of real symmetric matrices $Sym(n,\mathbb{R})$ that is naturally endowed by a metric structure using Frobenius inner product $\langle A, B \rangle = Tr(A \cdot B)$.

A simple network, with no self-loops, belongs to the set of traceless matrices $Sym_0(n,\mathbb{R})$. It is worth noting that while considering $A_{ij}$ as a real-valued link, we do not attempt here to give a criterion to quantify the strength of, say, the alliance between two Nations, or conversely their degree of foe (they could be possibly estimated on the grounds of commercial treats or diplomatic exchanges). All we assume is that friendship\foe between any two actors can be represented by a single (signed) real number.

## III   WALKS ON SIGNED NETWORKS

To begin with, we introduce a set of $n$ scalar functions

$$\phi_k(A) = Tr(A^k) \qquad k = 1, 2, \ldots n \qquad (3)$$

and generalize expression (1) defining potential energies

$$u_k(A) = -\frac{\phi_k(A)}{\phi_k(|A|)}. \qquad (4)$$

We recall that $\phi_0(A) = n$ represents the number of nodes of the network and $\phi_1(A)$ gives information on the presence of self-loops. The main rationale to generalize expression (1) is that, in a generic network, testing all triangles may not give an acceptable measure of global balance, as shown in [10]; consequently, normalized functions (4) quantify the balance of closed weighted k-polygons on the graph. More generally, by definition given in Ref. [11], a walk on a signed graph is positive iff it contains an even number of negative links, otherwise it is negative;. therefore the k-th power of adjacency matrix has $(A^k)_{ij}$ equal to the number of positive walks of length k from i to j, minus the number of negative walks, In other words in graphs containing negative edges, the powers of adjacency matrix represent signed path counts, where paths with an odd number of negative edges are counted negatively, thus implementing *enemy-of-an-enemy* multiplication rule generalized to arbitrarily long paths [12].

In order to give a physical meaning to negative walks, we recall the concept of communicability matrix [13],

$$G(A) = \exp(A) \qquad (5)$$

for a generic *unsigned* complex network (for a review on the argument see [14]). Communicability function (5) between nodes i and j essentially quantifies, within the framework of information theory, how long it takes to send a message from i to j; moreover its eigenvalues give a measure of nodes centrality in the network. Their sum, known also as Estrada index [15],

$$EE(G) = Tr(\exp(A)) \qquad (6)$$

is a topological index of the corresponding graph [16].

For general signed networks, we define as communicability matrix the exponential of its absolute adjacency matrix. $G(A) = \exp(|A|)$

Going back to antagonistic interactions in social networks, suppose to have an exchange of messages among actors of the network. Every time that a message crosses a negative link (i.e. through an enemy node) it is altered in transit; otherwise it is transmitted in the right way. Incidentally, it is plausible to consider the same process in the framework of information theory. Suppose that a bit of information is transferred along the network, and that every time the carrier walks through a negative (unreliable) edge, a bit flip occurs.

A measure of the *integrity* of the exchanges is obtained by subtracting the number of paths $n_{ij}^+$ ( open or closed), through which the uncorrupted information arrives, from the number of paths $n_{ij}^-$ through which it would be altered. This value corresponds to expression (5) applied to the signed adjacency matrix of the network. The normalized value of it $I_{ij} = \left( \dfrac{n_{ij}^+ - n_{ij}^-}{n_{ij}^+ + n_{ij}^-} \right)$ is expressed by Hadamard product of the matrix exponent (5) and the Hadamard inverse of its communicability matrix [17]

$$I(A) = \exp(A) \circ \exp(|A|)^{\circ(-1)} \qquad (6).$$

For all the possible paths from i to j, signs of $I_{ij}$ represent the reliability/unreliability associate to information transfer among nodes. Module of its entries can be taken as a measure of the difference between the two probabilities to choose a reliable/unreliable walk inside this set. Values around zero $I_{ij} \approx 0$ are associated to sets of paths characterized by maximum uncertainty, meaning equiprobability to choose a positive or negative walk.

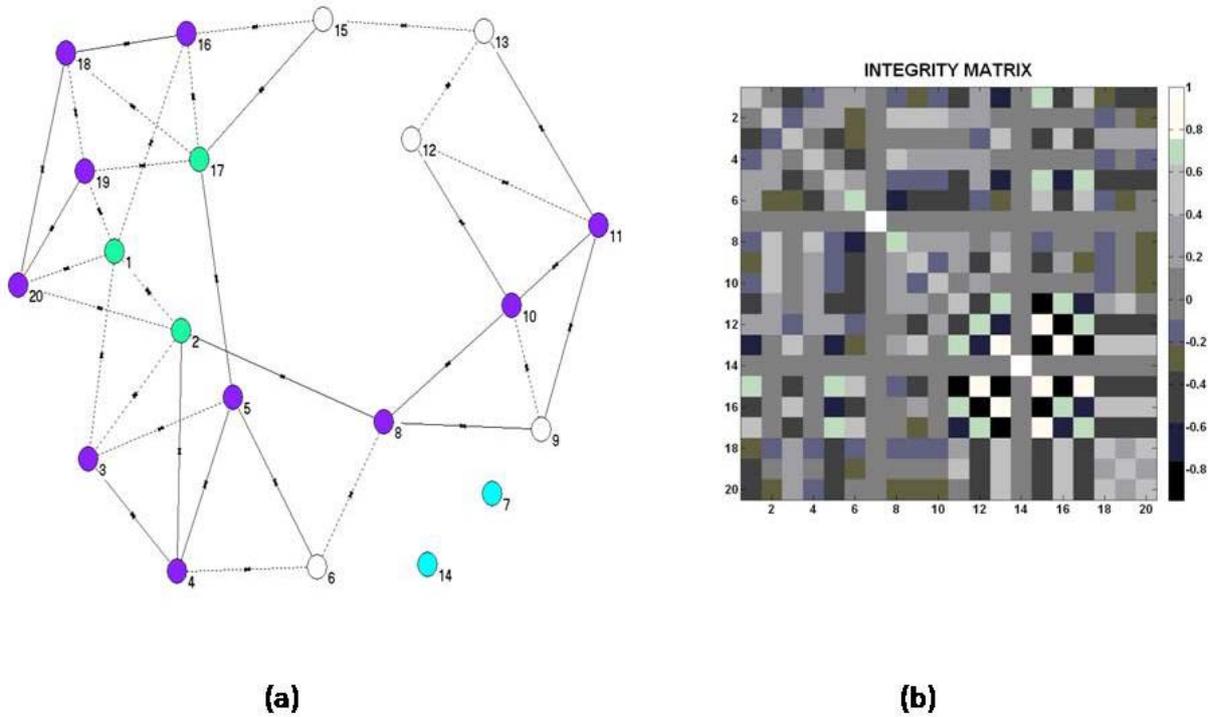

**FIG.1**

**FIG.1** (a) a signed network having 20 nodes and its relative Integrity matrix. (b).Nodes 14 and 7 are isolated $I_{jk} = \delta_{jk}$ $(j = 7,14 \quad k = 1,2...20)$. While there is a high probability to make a negative walk from nodes 1 to 13 $I_{(1,13)} = -0.7$, on the contrary a walk between nodes 1 and 15 has god chances to be positive, $I_{(1,15)} = 0.7$. Finally a walk between nodes 1 and 19, $I_{(9,19)} = 0.1$ has both chances to be negative or positive.

We point out that, because of Cartwright-Harary theorem, it is easy to demonstrate that in a balanced network, Integrity matrix $I_{ij}(A) \in Sym(n,\{-1,+1\})$; moreover we have $Tr(I) \leq n$, where equality holds for balanced networks, while $Tr(I)$ can be taken as a measure of balance.

At this point we are ready to define a dynamical system for a signed network.

## IV. ISOSPECTRAL FLOWS

As mentioned in Section II, full balance states are stable fixed points of the system that configures itself in a perfect polarization of the corresponding graph with $u_k(A) = -1$. In dynamical systems a minimum energy-state configuration is reached when a dissipation acts. As a matter of facts equation (2) can be easily related to the matrix valued function $\phi_3(A)$. Using $\frac{\partial Tr(F(A))}{\partial A} = f(A)^T$, where $f(\cdot)$ is the scalar derivative of $F(\cdot)$, because $A = A^T$ we can write the time evolution of the network potential energy (2), from which, assuming a natural gradient system dynamics

$$\dot{A} = -\frac{\partial}{\partial A} \phi_3(A) = \frac{9}{\binom{n}{3}} A^2 \qquad (7)$$

we recover equation (2), that was assumed as ansatz in Ref.[4]. Furthermore we recognize potential (1) as $u_3(A)$. Geometrically a balanced network is a specific point belonging to the manifold whose generic points are matrices of $Sym(n, \mathbb{R})$, in this $\frac{n(n+1)}{2}$ dimensional space it is possible to write a dynamical system whose evolution is conservative with respect to the structural properties of the initial network $A(t_0)$. The ordered set of $n$ eigenvalues of the adjacency matrix is called the spectrum of the graph, cospectral graphs, also known as isospectral graphs, are graphs that, although generally not isomorphic, share the same graph spectrum

It is possible to write a network dynamics that preserves the adjacency spectrum ,together with graph invariants as characteristic polynomial $\pi_G(A) = \det(A - \lambda I)$. This evolution is known as isospectral flow [18]

$$\dot{A} = [A, f(A)] \qquad (8)$$

given by a Lax pair, where $[A, f(A)] = A \cdot f(A) - f(A) \cdot A$ is the matrix commutator and $f(A) = -f(A)^T \in so(n)$ is a skew-symmetric matrix.

As regards functions (3), that written in eigenvalues form become $\phi_k(A) = \sum_{j=1}^{n} \lambda_j^k(A)$ [18], using trace invariance under cyclic permutations, we get the conservation property

$$\frac{d\phi_k(A)}{dt} = Tr\left(\frac{\partial \phi_k}{\partial A} \frac{dA}{dt}\right) = Tr\left(\frac{\partial \phi_k}{\partial A}[A, f(A)]\right) = 0. \qquad (9)$$

furthermore, because of isospectrality, Estrada index is also conserved

$$\dot{E}(G) = \frac{d}{dt}\left(\sum_k \exp(\lambda_k)\right) = 0. \qquad (10)$$

Equation (8) describes a social dynamics where friendship between actors continually evolves but never collapses into a pair of stable factions. As regards to signed Laplacian matrix $L_{ij} = \bar{D}_{ij} - A_{ij}$, where $\bar{D} = Diag(\bar{d}_1, \bar{d}_2, ..., \bar{d}_n)$ and $\bar{d}_i = \sum_{k=1}^{n}|A_{ik}|$ is the signed degree matrix [19], the Laplace spectrum is not conserved. Moreover, its minimum eigenvalue known as algebraic conflict, $\xi$, is always positive for the dynamics described by eq.(8). (Fig.2). It represents an index of conflict in the network, whose value is exactly zero when $G$ is balanced [8].

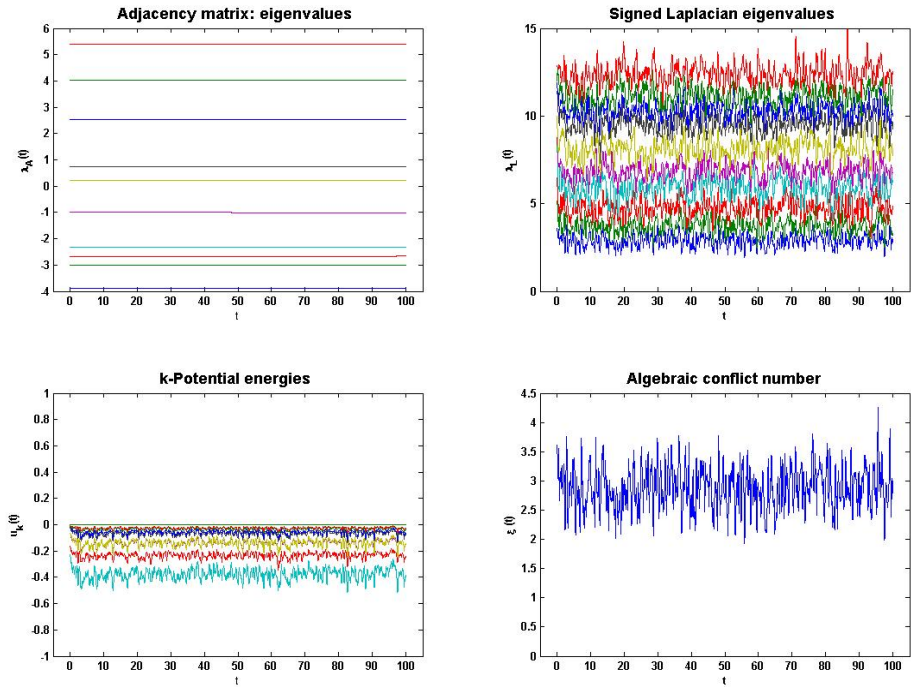

**FIG.2**

**FIG.2.** Free dynamics of a 10 nodes random network. Here self-loops are allowed for the evolution, choosing an unconstrained random skew symmetric matrix. Potentials and algebraic conflict maintain their values far from a stability network configuration.

In order to check numerically the variability of $A_{ij}(t)$, in Fig.3 we represent the triangular matrix $\langle A_{ij} \rangle$ for $i \leq j$, whose $\dfrac{n \cdot (n+1)}{2}$ non-zero entries are the time average of link strengths.

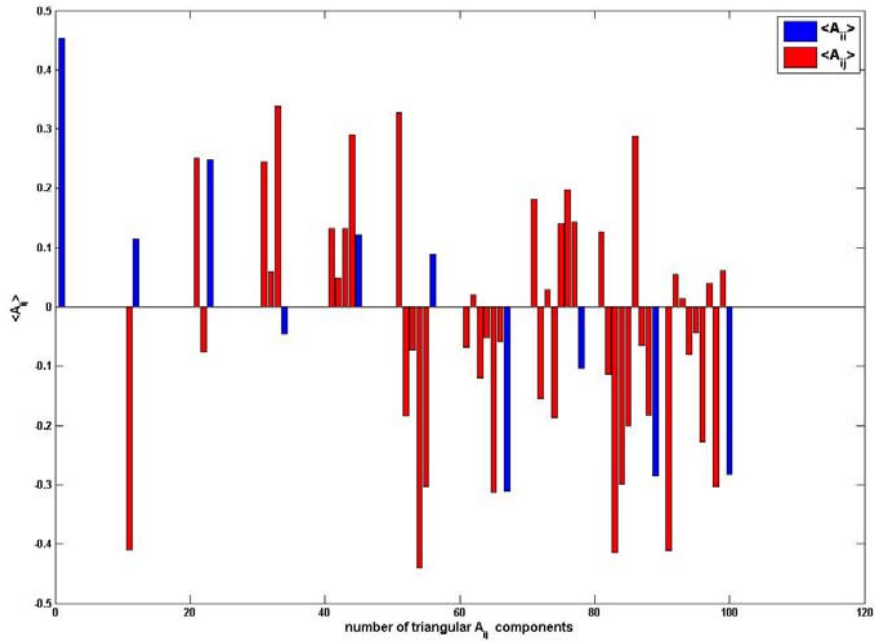

**FIG.3**

**FIG.3.** Average values of $\tilde{A}_{ij}$ entries. (Colors) Blue bars represent self-loops, red ones off-diagonal links. It can be noted that they are numerically comparable.

Besides, using a fast method to extract community structures [20], this kind of dynamics shows weak and transitory small communities (from 2 to 4 nodes) with a modularity value around 0.4 . As regards to instability, trace of the mean integrity matrix (6) gives a value very far from stability ($Tr(\langle I \rangle) \sim 1$). Images of the corresponding graphs at four different times are shown in Fig.4.

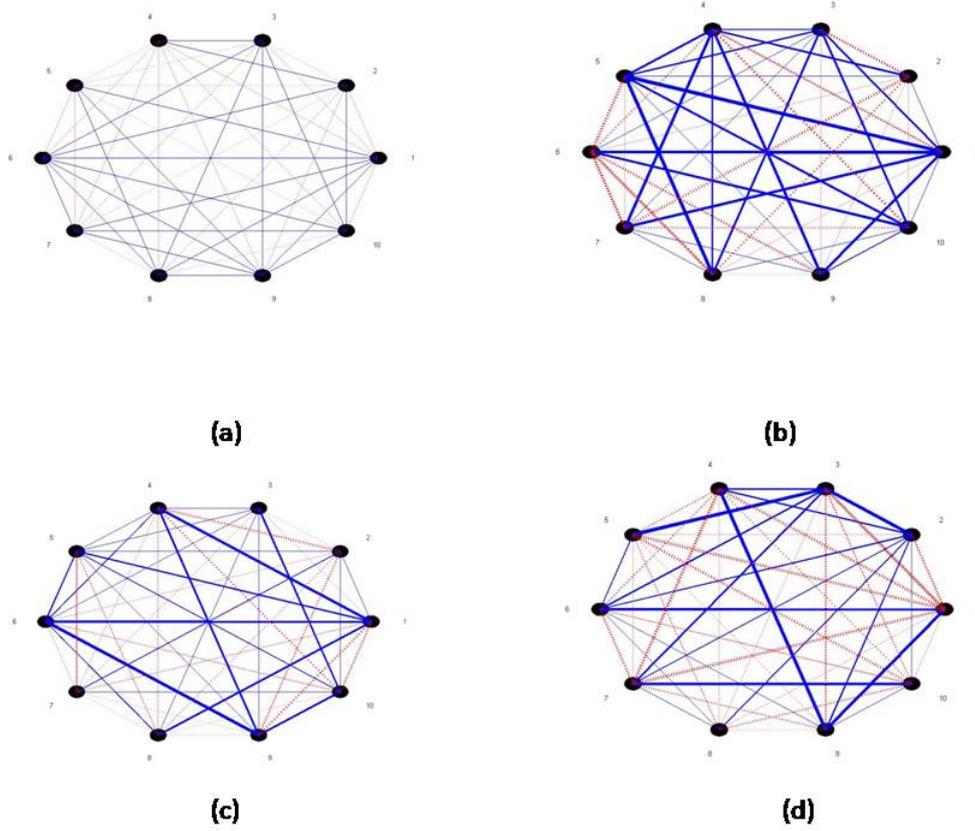

**FIG.4**

**FIG4.** Cospectral Graphs associated to equation (9). Pictures taken at initial condition (a) and after $k \cdot \frac{t_{final}}{4}$ $(k=1,2,3)$ steps. Initial adjacency matrix: $A(t_0) \in Sym_0(10, \mathbb{Z}_2)$, $diag(A(t_0))=0$. (Colors) Blue continuous lines represent positive links, Red dashed lines negative links. Link thickness represent $|A_{ij}|$, self-loops are not shown.

Choice of function $f(A)$ constraints more elements of the net to be preserved; for instance an initially isolated node $s_K$ $(A_{Kj}(0)=0 \ \forall s_j \in V)$ will remain isolated $(A_{Kj}(t)=0, t>0)$ under the choice $f(A)_{Kj}=0 \ \forall s_j \in V$. For simple networks, while $Tr(A)$ is an invariant, $Diag(A)$ is not generally conserved and, as for equation (2), during the evolution self-.loops appear. In order to avoid them, a constraint must be imposed by lowering the degrees of freedom of $f(A)$ from $\frac{n(n-1)}{2}$ to $\frac{(n-1)(n-2)}{2}$ free entries. Adjacency matrix is given by solving the system

$$\begin{cases} \dot{A} = [A, f(A)] \\ Diag[A(t), f(A(t))] = 0 \quad \forall t \\ A(t_0) \in Sym_0(n, \mathbb{R}) \end{cases} \quad (11)$$

and constrained skew symmetric operator will assume the following form:

$$f(A)_{ij} = \begin{cases} 0, & i = j \\ A_{ij}^{-1} \cdot \sum_{k=2}^{n} S_{jk} A_{jk}, & i = 1, j > 1 \\ S_{ij}, & i \neq 1, i < j \end{cases} \quad S \in so\left(\frac{(n-1)(n-2)}{2}, \mathbb{R}\right) \quad (12)$$

For a simple 5-nodes network a simulation is shown in Fig.5 using a second order Heun integrator.

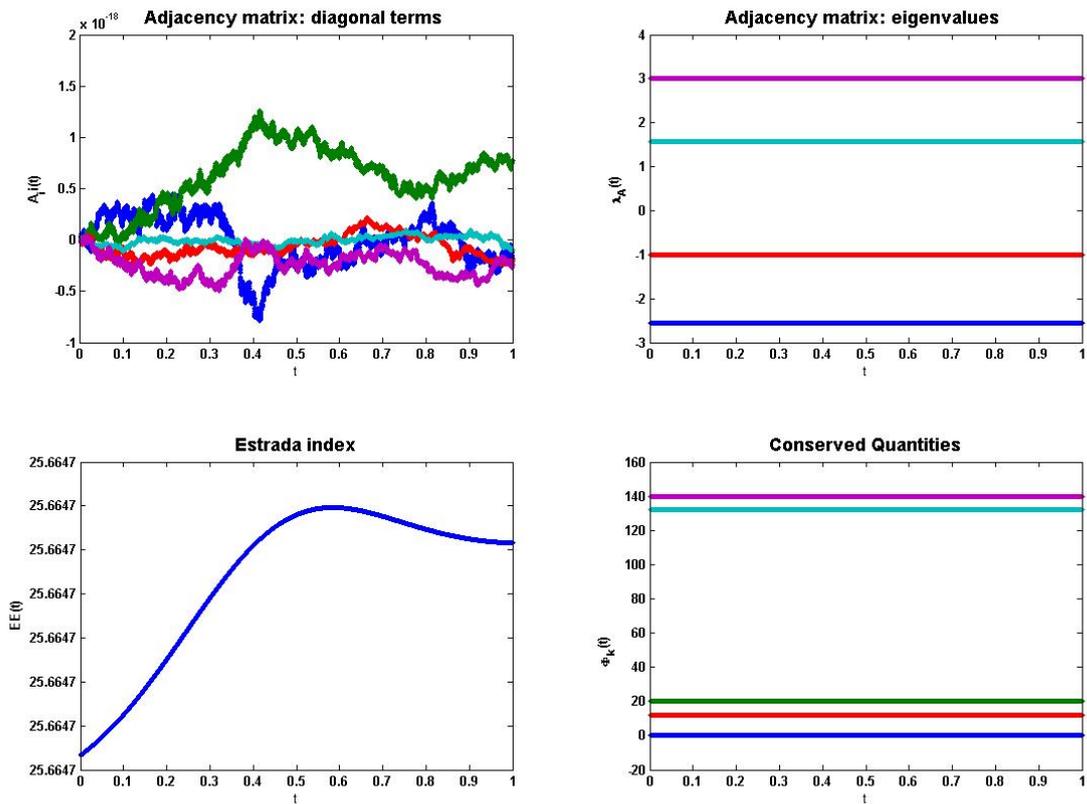

**FIG.5**

**FIG.5** A simulation of the constrained simple network as in equations (12) where self-loops do not rise.

Because of the geometric nature of Lax equations (8), a symplectic integrator would better perform longer time integrations. Free dynamics allows the net to explore a large part of $Sym(n,\mathbb{R})$ manifold and, because of conserved quantities, at the same time its trajectory is constrained to stay on the submanifold represented by (9).

## V. LINK-POTENTIALS: JAMMED AND BALANCED STATES

In order to analyze more interesting behaviors of (9), we imagine $\phi_2(A)$ as a sort of "kinetic energy" of the signed network and introduce a set of dissipative potentials given by polynomials s $P_A(\phi) = \sum_k \alpha_k \cdot \phi_k(A)$, where $\alpha_k \in \mathbb{R}$. In this way the evolution of the system will be given by equation

$$\dot{A} = [A, f(A)] + \frac{\partial}{\partial A} \cdot P_A(\phi) \qquad (13)$$

giving rise to a gradient system mechanism. In case $P_A(\phi) = \phi_2(A)$, we found a pure dissipative system where net evolves to a completely disconnected set. In a language proper to social networks, we could imagine this configuration as a final state of actors who become *lurkers* because of lack of participation [21]; from a geopolitical point of view, it could be translated into the word *isolationism*.

For a certain choice of the polynomial, such as $P_A(\phi) = \phi_3(A) - \alpha \cdot \phi_4(A)$, dynamics (13) brings network into a jammed state; indeed simulations show that values of $u_k(A)$ are close to -1 and values of algebraic conflict are close to zero, meaning a more balanced configuration with respect to free dynamics (9). (Fig.6)

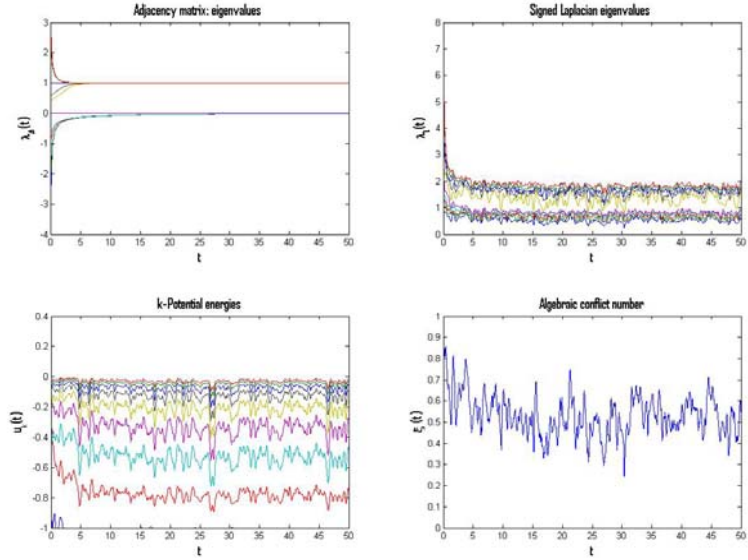

**FIG.6**

**FIG.6.** dynamics of a 10 nodes random network within potential $P_A(\phi) = \phi_3(A) - \phi_4(A)$. Self-loops are allowed for the evolution, choosing an unconstrained random skew symmetric matrix.- Potentials and algebraic conflict maintain their values far from a balanced network configuration but in jammed state closer to .-1. See Fig.2 for comparisons.

Concerning global balance, we have seen that using formalism (2) a balance state is reached with link strengths going to infinity. Considering numerical values of links as measures of friendship/foe, looking at social interactions, or similarly at international relations, it is natural to imagine that some alliances/hostilities can be stronger than others; however, a complete annihilation of national or self-identity is unrealistic. Then, it follows that every person/Nation does impose a certain natural limit to friendship/alliances with other entities. As an example, let's consider sensitive information which might endanger national security, or people who have a deeply personal secret they will never share with anyone. Here we assume that every actor in the net has a proper boundary that no-one can penetrate, and we describe this fact introducing a single-link potential in our formalism, in addition to multi-link potentials (3). Above a certain link strength it will act in a repulsive way, thus avoiding unphysical infinities. This mechanism is easily provided by the Hadamard power $A^{\circ(k)}$ of the adjacency matrix. For $P_A(\phi) = \phi_3(A)$, equation

$$\dot{A} = \left[ A, f(A) \right] + A^2 - \alpha A^{\circ(k)} \qquad (14)$$

thus brings the net to a balanced state as in ref. [6], but avoiding infinities in the link strength. In Fig. 7 final balanced state for equation (14) with finite link strengths is shown for a Watts-Strogatz type initial condition.

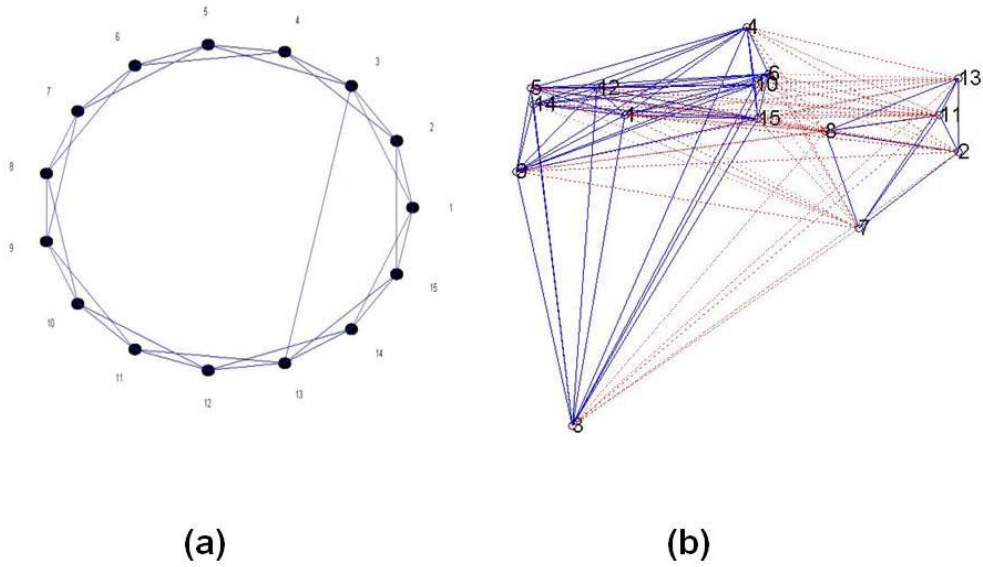

(a)  (b)

**FIG.7**

**FIG.7** Evolution to a balanced state for a 15 nodes small-world network using eqns . $P_A(\phi) = \phi_3(A) - \left[ \dfrac{\partial \phi_2(A)}{\partial A} \right]^{\circ(4)}$ . Fig(a) initial configuration, (b) final graph with 2 opposed factions. Final link weights of the normalized Adjacency matrix $N_{ij}(t_{final})$ fall inside an interval $[0.92, 1]$.

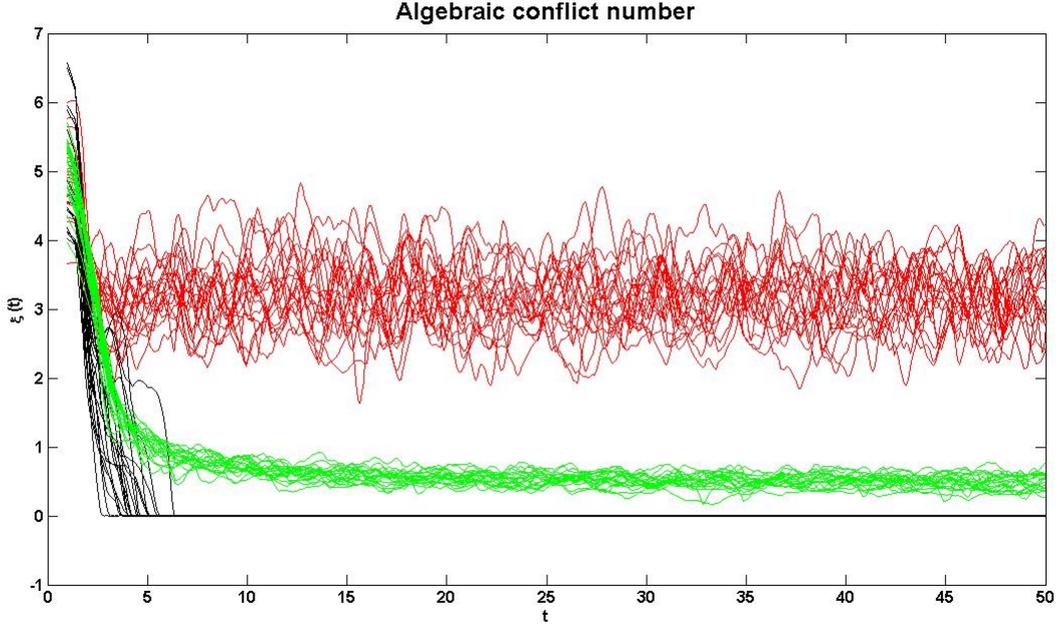

**FIG.8**

**FIG.8.** Time evolution of algebraic conflict number for a number of random complete 20 nodes signed networks with $A(t_0) \in Sym_0(20, \mathbb{Z}_2)$, $diag(A(t_0)) = 0$ for different choices of potentials: (Colors), Red $P_A(\phi) = 0$, Green, $P_A(\phi) = \phi_3(A) - \phi_4(A)$, Black $P_A(\phi) = \phi_3(A) - \left[\frac{\partial \phi_2(A)}{\partial A}\right]^{o(5)}$.

An interesting dynamics has been found taking as initial condition a balance state; for a free dynamics, of course, we find no evolution under condition (12). Allowing for self-loops to evolve in time under eqn. (8), the sign oscillations of diagonal terms of adjacency matrix make the net unstable giving rise to 'internal conflicts' (Fig.9). We point out that even if the network shows a balance between nodes, self-loops alternating behavior rise the algebraic conflict number (Fig.10).

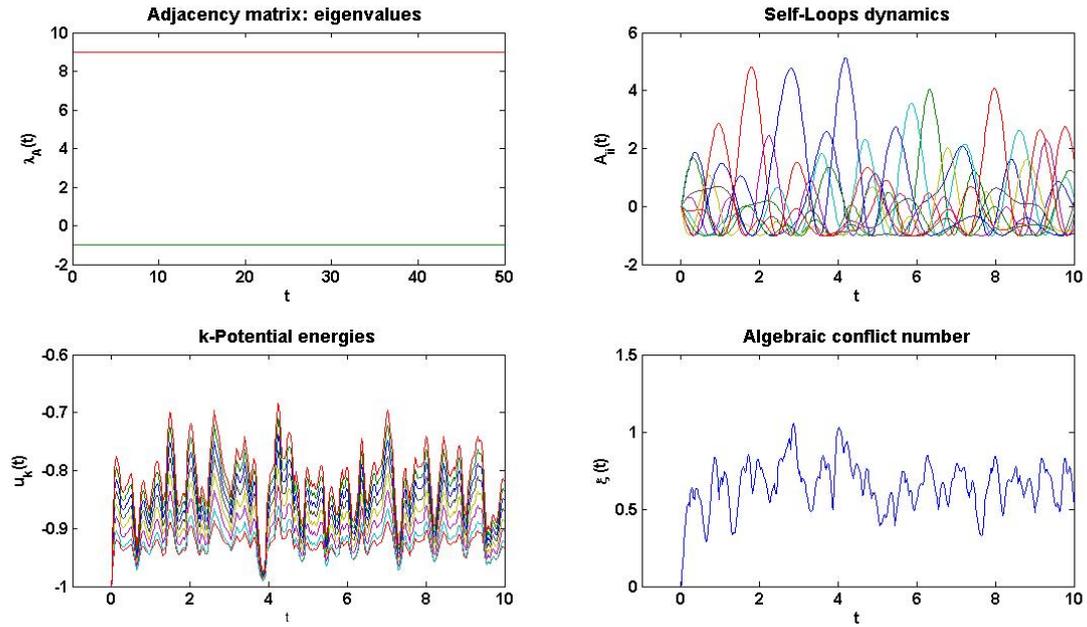

**FIG.9**

**FIG.9.** Self-loops effects on an isospectral evolution (9). for a simple net with eigenvalues $\lambda_A^i = -1, (i = 1...9)$, $\lambda_A^{10} = 9$ and starting in a balanced state. Oscillating values of Adjacency matrix create net unbalance because of internal conflicts of actors.

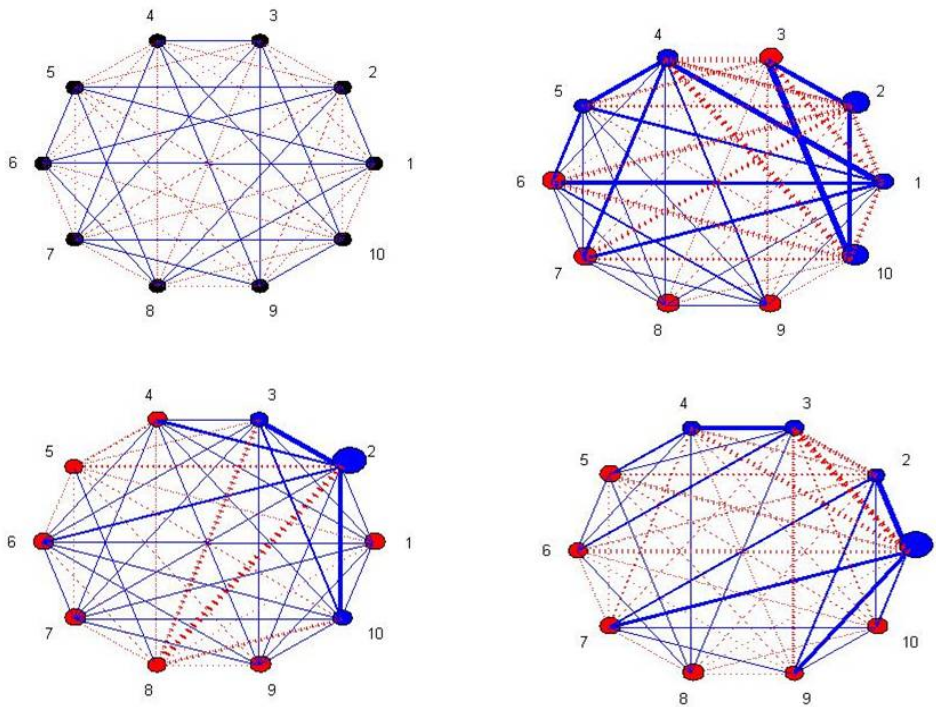

**Fig.10**

**FIG.10** Cospectral graphs associated to eqn.(9) and starting from a balanced state. Red colors show negative links/loops, blue colors positive links/loops. Thickness is proportional to their numerical values. Note the triad balance.

## VI PERTURBING TRIADIC RELATIONSHIPS

From a general point of view, in dealing with social balance potential, we recognize that at least one important element relevant for conflict dynamics is lacking in our modeling. To be specific, the very notion of stability and instability of a triad embedded in a complex network, depends not only on the interplay of alliances within the triad itself, but also on the remaining nodes.

Consider the situation depicted in Fig.11, where triad $(i,j,k)$ shows an unstable state; structural balance theory then predicts that such a configuration evolves (preferably) towards balance by a change in sign of $A_{ij}$ or $A_{kj}$. Suppose now that node $s_j$ is almost positively linked to other nodes, external to the original triad; it is then easy to imagine that, in spite of the existing instability, it is more difficult for actor $s_i$ to break alliance with actor $s_j$. At the same time actor $s_k$ would consider more convenient to lower his hostility with $s_j$ (at least in the situation in which $s_k$ has not as strong alliances as $s_j$).

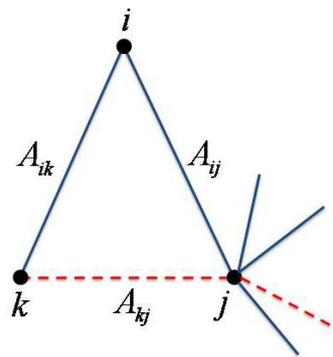

**FIG.11**

**FIG.11.** Triadic relationships affected by a node having a high degree of alliances.

More quantitatively, a simple small correction to balance potential $\phi_3(A)$, can be introduced using network degree matrix values $d_i = \sum_{k=1}^{n} A_{ik}$,

$$\tilde{\phi}_3^{(\gamma)} = Tr(\tilde{A}^3) \tag{15}$$

where $\tilde{A}_{ij} \equiv A_{ij}(1 + \gamma \cdot |d_i - d_j|)$, $\gamma > 0$. Equations assume (in explicit matrix elements representation) the following structure :

$$\dot{A}_{ij} = [A, f(A)]_{ij} + (\tilde{A}^2)_{ij} \cdot (1 + \gamma|d_i - d_j|) + 2\gamma \sum_l A_{lj}(\tilde{A}^2)_{lj} \cdot sign\left(\sum_h (A_{jh} - A_{lh})\right) \tag{16}$$

As an example of it, we illustrate in Fig.12 two different dynamics driven by potential (15) where both initial networks have node $s_1$ as hub but with opposite degree. Final nets show that this node keeps its property of super allied/enemy, increasing its degree. In the first case a whole community of friends comes out; in the second one, node $s_1$ becomes enemy of all others.

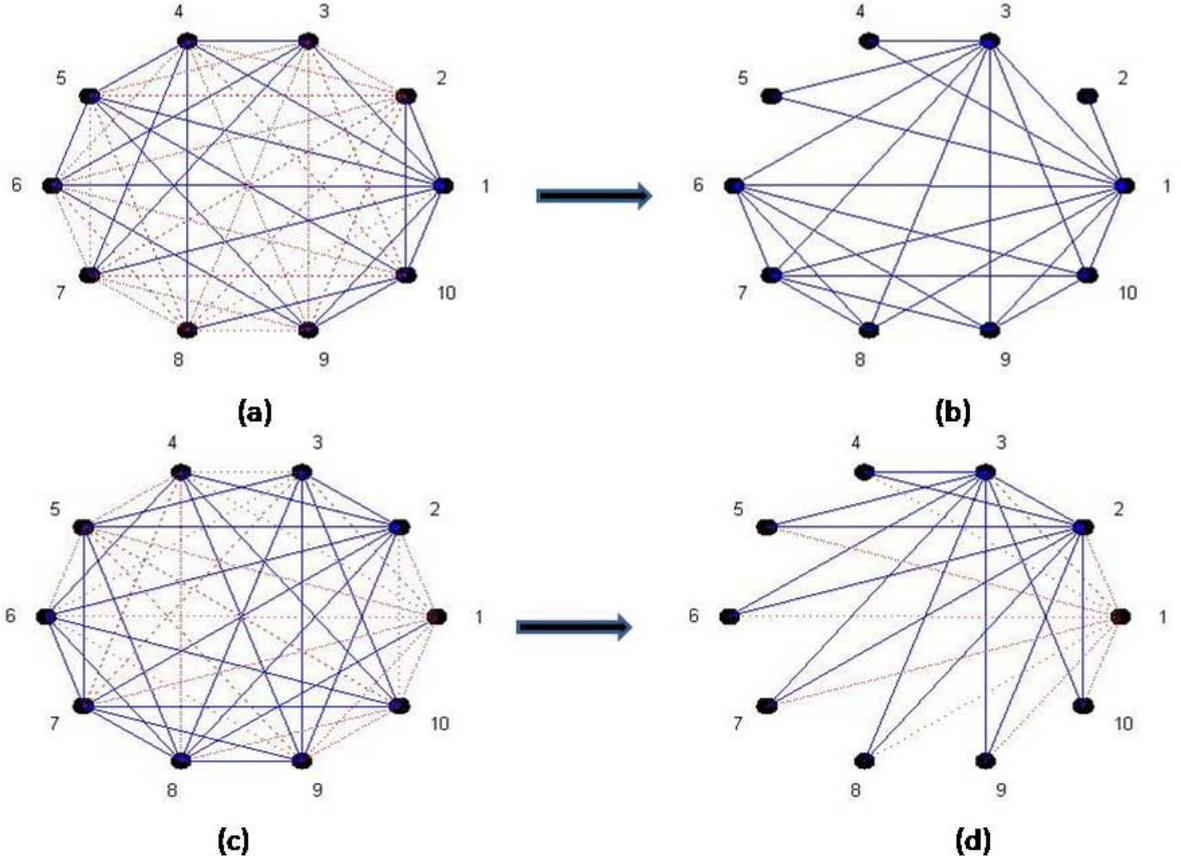

**FIG.12**

**FIG.12.** Effects due to potential links corrections. Two initial and link-opposite networks (a) and (c), (both possessing a hub in node 1) are driven by eqn.(16). In final nets (b) and (d) the same hub has more allies/enemies then other actors. A threshold $\tau = \max\left(\|A(t_{final})\|\right) \cdot 0.05$ has been used in order to cancel minor links.

## VII    A GENERAL CLASS OF DYNAMICAL MODELS

In this section our aim is to define a general class of dynamical models which includes the previously analyzed behaviors as special cases. Our purpose is twofold: 1) to describe possible heuristic models with free parameters to be estimated form real complex networks, where conflict dynamics, as in a social network, is present; 2) to obtain a class of toy-models useful to simulate realistic evolving conflicts. The general equation of this class is

$$\dot{A} = [A, f(A)] - \sum_{k>2} \alpha_k \frac{\partial}{\partial A} \tilde{\phi}_k^{(\gamma)} + \sum_{k\geq 0} \beta_k A^{\circ(k)} \qquad (17)$$

where $\tilde{\phi}_k^{(\gamma)} = Tr(\tilde{A}^k)$.

Here we give a summary of the various terms on the r.h.s. of (17):

1) Isospectral term represents the free dynamics (seemingly ergodic on the manifold at fixed trace invariants) which does not give variations of social potentials leading to balanced or quasi-balanced (jammed) states. It represents a random link evolution of non-polarizing formation and variation of conflicts and dynamics. A physical analogy can be found in ferromagnetic materials where, in absence of an external magnetic field, spins evolve randomly with no statistically preferred but sporadic configurations;

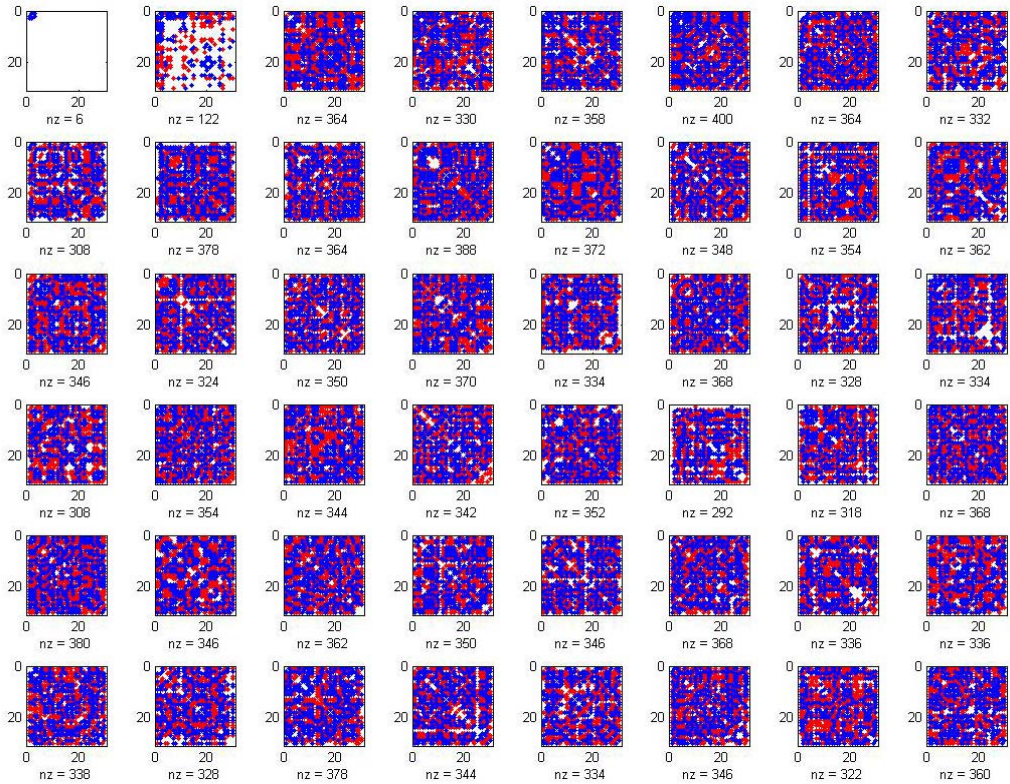

**FIG.13**

**FIG.13.** A representation of adjacency matrix for an evolving net of three initial nodes for different time steps. Here only isospectral term is involved in eqn (17). Red dots represents negative links, while blue dots positive ones. A threshold $\tau = \max\left(\left|A\left(t_{final}\right)\right|\right) \cdot 0.05$ has been used in order to cancel minor links. A reordering algorithm has been applied in order to visualize possible sporadic communities. nz counts the number of red dots.

2) Social balance term represents the tendency of the network to cluster into factions based on triadic or polygonal closed loops; $\gamma$-multiplied term provides for a correction for this tendency accounting for the effect of nodes degree( or alliance degree), eventually amplifying or depressing the tendency towards local balance. This last effect is relevant for the emergent features in evolutionary networks.

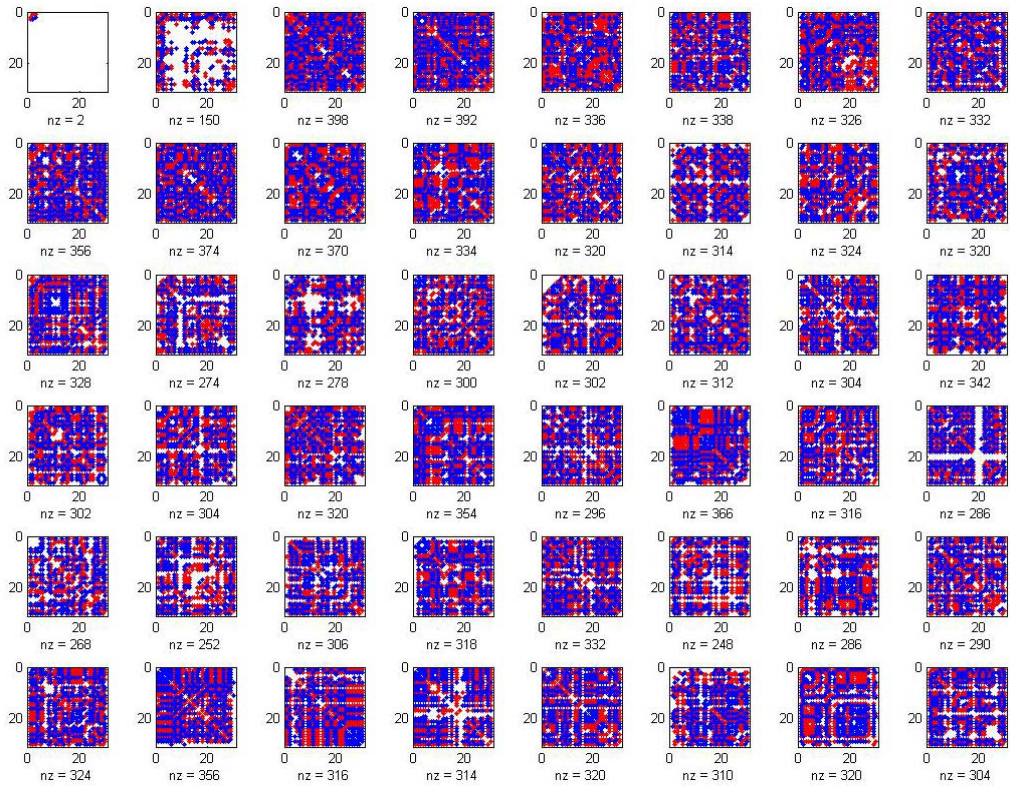

**FIG.14**

**FIG.14.** A representation of adjacency matrix for an evolving net of three initial nodes for different time steps. Here both isospectral and second term is involved in eqn (17). Red dots represents

negative links, while blue dots positive ones. A threshold $\tau = \max\left(\left|A\left(t_{final}\right)\right|\right) \cdot 0.05$ has been used in order to cancel minor links. A reordering algorithm has been applied in order to visualize possible communities, nz counts the number of red dots.

3) Finally single link potential represent the resistance of the actor, or privacy barrier, eventually avoiding alliances/hostilities to indefinitely grow.

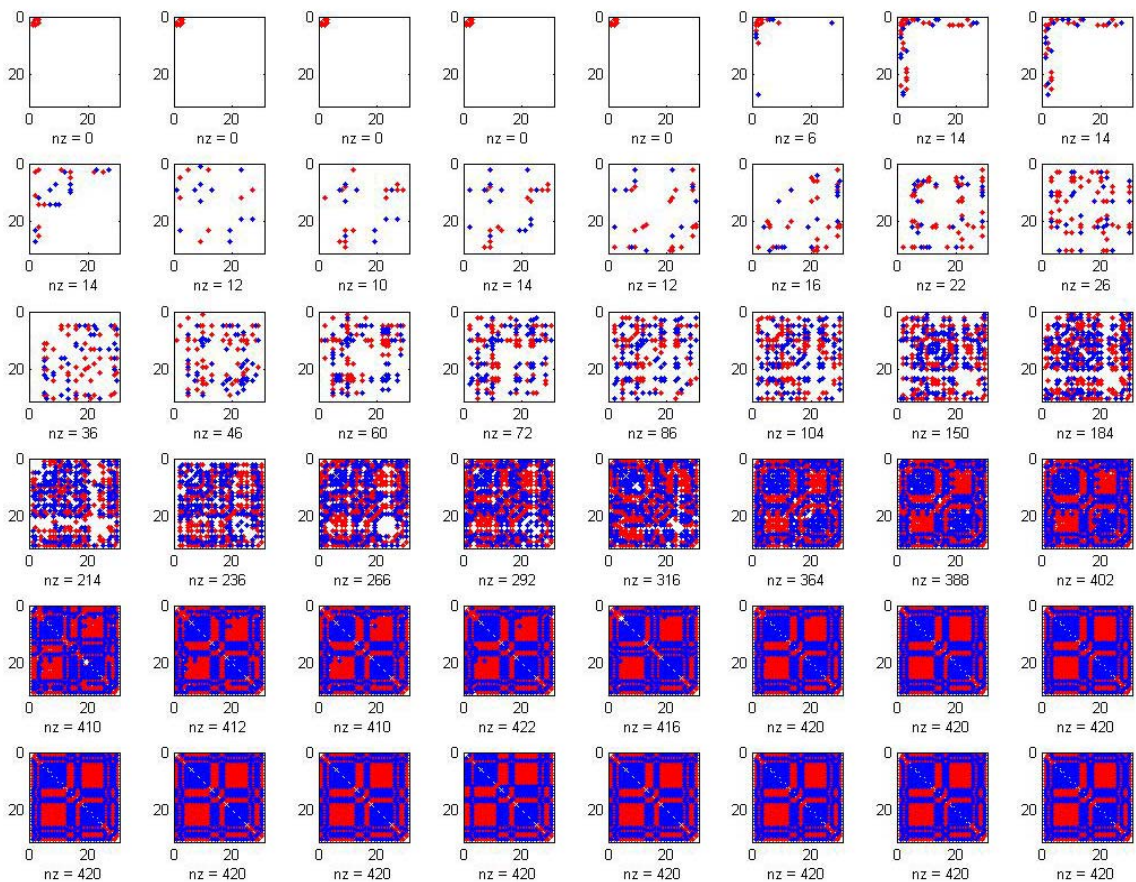

**FIG.15**

**FIG.15.** A representation of adjacency matrix for an evolving net of three initial nodes for different time steps. Here all terms involved in eqn (17) are in action. Red dots represents negative links, while blue dots positive ones. A threshold $\tau = \max\left(\left|A\left(t_{final}\right)\right|\right) \cdot 0.05$ has been used in order to cancel minor links. A reordering algorithm has been applied in order to visualize growing communities and final factions, nz counts the number of red dots.

Coming back to the possible use of (17) as an empirical model with free parameters, a minimalistic such a model, with a relatively small number of parameters, can be written as

$$\dot{A} = [A, f(A)] - \alpha \frac{\partial}{\partial A} \tilde{\phi}_3^{(\gamma)} + \beta A^{\circ(2n+1)} \qquad (18)$$

where $f(A)$ contains only one free parameter, corresponding to taking in eqn (12) $S_{ij} \equiv s$, and $n \geq 1$ is an integer. Total number of free parameters is then five: $(s, \alpha, \gamma, \beta, n)$.

**VIII CONCLUSIONS**

Studies on digital society from a mathematical viewpoint, are receiving increasing interest in contemporary scientific research [22]. In this area, conflict dynamics described by a dynamic change of the adjacency matrix of a signed complex network, gives opportunities to explore dynamical systems on symmetric matrices space $Sym(n, \mathbb{R})$. Free dynamics, assumed as isospectral flow, conserves a set of n trace-invariants and gives (pseudo) random formation/disruption of alliances without any definite tendencies towards social balance.

Jammed states, similar to those described in signed network discrete-time evolution, are found by inserting appropriate scalar potentials of the same form of invariants, showing a low but non-zero algebraic conflict number. Furthermore, structural balance theorem is recovered introducing a gradient system and appropriate single link potentials given by the Hadamard powers of the adjacency matrix.

Besides, the model gives the opportunity to consider the strength of links in the graph, associating numerical indices to enemy/friendship relations among network actors.

Finally, a general class of models for a signed network as a dynamical system has been proposed within a self-consistent architecture of the unifying formalism based on trace-invariants of the free isospectral flow. Applications to real networks will be discussed in a forthcoming paper.

**ACKNOWLEDGEMENTS.**

The authors acknowledge the interesting e-mail exchanges with T. Zaslavsky on the argument negative walks in signed networks. This work is part of OSN-CASD project "Cyber World" and CEMISS research project on Complexity.